\documentclass[aps,prb,showpacs,floatfix,amsmath,amssymb,
               preprint]{revtex4-1}
\usepackage{color}
\usepackage{graphicx}
\usepackage{natbib}
\usepackage{epsfig}
\usepackage{amsmath}
\usepackage{amssymb}
\usepackage{verbatim}
\usepackage{bibentry}
\usepackage{indentfirst}
\usepackage{titlesec}
\titlespacing*{\section}
{0pt}{4.5ex plus 1ex minus .2ex}{4.3ex plus .2ex}
\titlespacing*{\subsection}
{0pt}{1.5ex plus 0.5ex minus .2ex}{4.3ex plus .2ex}

\begin{document}
\title{$Z_{2}$ Topological Order and Topological Protection of Majorana Fermion Qubits}
\author{Rukhsan Ul Haq }
\affiliation{Theoretical Sciences Unit \\ Jawaharlal Nehru Center for Advanced \\Scientific Research Bangalore, India}
\author{Louis H. Kauffman}
\affiliation{ Department of Mathematics, Statistics and Computer Science,  University of Illinois at Chicago,  \mbox{851 South Morgan Street}, Chicago, IL 60607-7045, USA; kauffman@uic.edu; and
	Department of Mechanics and Mathematics,
	Novosibirsk State University,
	Novosibirsk
	Russia.}
\begin{abstract}
The Kitaev chain model exhibits topological order that manifests as topological degeneracy, Majorana edge modes and $Z_{2}$ topological invariance of the abulk spectrum.  This model can be obtained from a transverse field Ising model(TFIM) using the Jordan-Wigner transformation. TFIM has neither topological degeneracy nor any edge modes. Topological degeneracy associated with topological order is central to topological quantum computation. In this paper we will explore topological protection of the ground state manifold in the case of Majorana fermion models which exhibit $Z_{2}$ topological order. We will show that there are at least two different ways to understand this topological protection of Majorana fermion qubits: one way is based on fermionic mode operators and the other is based on anti-commuting symmetry operators. We will also show how these two different ways are related to each other. We provide a very general approach of understanding the topological protection of Majorana fermion qubits in the case of lattice Hamiltonians.
\end{abstract}
\maketitle







	
\section{Introduction}
	
Topological order has recently emerged as a new paradigm for understanding topological phase of matter, which can not be captured by Landau's spontaneous symmetry breaking theory.One of the important manifestations of topologial order is the topological degeneracy of the ground state manifold. This degeneracy is topological in the sense that no local perturbation can lift it and hence there is topological protection of the ground state manifold. Topological degeneracy gives rise to a topological qubit for topological quantum computation. In this paper we will explore the protection of topological qubits in the context of Majorana fermions.  Majorana fermions have recently emerged as promising candidates for topological quantum computation. Consequently, Majorana Fermions have been on the research frontiers in condensed matter physics, and especially in topological quantum computation. 
One main mathematical difference between Majorana fermions and standard fermions is that the latter satisfy Grassmann algebra while the operator algebra of Majorana Fermions is a Clifford algebra. Clifford algebra of Majorana fermions attributes them with anyonic statistics - which is central for their role in topological quantum computation. The Clifford Braiding Theorem \cite{Kauffman2} gives a rigorous way of understanding this relation of Clifford algebra and anyonic statistics.


	
In \cite{Kitaev}, Kitaev introduced a quadratic Hamiltonian for fermions in one dimension, which has a topological phase in which there are Majorana modes at the edges of the chain. Kitaev employed a Majorana fermion representation to diagonalize the Hamiltonian and showed that there are Majorana edge modes. The Kitaev model is not an entirely new model. It can be obtained from the transverse field Ising model(TFIM) using a Jordan-Wigner transformation. So what Kitaev did is to take fermions as degrees of freedom instead of spins. This novel point of view opened up the way to understand topological order in a well-known system. TFIM exhibits only Landau order, and the ordered phase arises due to the symmetry breaking of the model. The immediate question that comes to mind is how is Landau order in TFIM related to the topological order, or how does topological order arise in the Kitaev Chain when one maps to a fermionic representation. 
The relation between these two models has been studied in a recent work\cite{Greiter}. There it is concluded that spectral properties of the two models are same, which we will find below is not correct. Using dualities and a bond-algebra approach and holographic symmetries, there is already an understanding of how topological order in the Kitaev chain model is related to Landau order in the corresponding spin model \cite{Ortiz}. Under the duality transformation, local variables of spin models get transformed to non-local observables which are important for topological order in the fermionic model. The work  \cite{Ortiz} also shows that Hamiltonians that exhibit topological order have holographic symmetries. In an another work \cite{Fendley} Fendley has found an algebraic approach for topological order in a Majorana chain, and more generally for a parafermion chain. In Fendley's paper, a fermionic mode operator is defined and it is shown that in the topological ordered phase, there are these fermionic mode operators that  for the Kitaev chain model are the Majoarana mode operators. As with Fendley, we shall also find that in the topologically ordered case, there are emergent symmetry operators that are actually Majorana mode operators. 
	
Normalized zero modes have attracted a lot of attention recently not only in topological phases of matter but also in many-body localization(MBL) where they offer a very intuitive language for understanding MBL phenomenology\cite{Fendley,Mier,Fend2,Nand,Altman,Param}. Normalized zero modes can be even or odd depending on whether they commute or anti-commute with the Hamiltonian for the quantum system. Even zero modes are also called pseudo-spins in MBL literature. In this paper, we take an approach of understanding topological protection of Majorana fermions qubits based on Majorana zero mode operators which are odd normalized zero modes. Similar to even mode operators in MBL, we show that odd fermionic operators are also integrals of motion and the emergent symmetry operators of the Hamiltonian. Our approach brings out new aspects of the topological order in Majorana fermion models especially the connection to emergent symmetry operators.

	
	
The rest of this paper is organized as follows: First, we briefly discuss the Kitaev chain model, its symmetries, and topological order. Then in the next section, we show how the algebra of Majorana fermions is conceptually incomplete, and we introduce emergent Majorana fermions. In section 3 we show how supersymmetry arises in Majorana fermion models. In Section 4, we show that the operator for emergent Majorana fermion is actually a Majorana mode operator. We also generalize our results for an odd number of Majorana fermions both without interaction as well with quartic interactions. We briefly talk about effects of disorder on topological order in Majorana fermion models.In section 5, we explore the relation between topological order and the braiding representation of Majorana fermions. We show that this braid group representation has extra symmetries where the Majorana mode operators are the symmetry generators. Finally we summarize our results in section 6.
	
\section{Kitaev p-wave  Chain}
To study the relation between Landau order and topological order we introduce two Hamiltonians that are related to each other by a Jordan-Wigner transformation. The two models are a transverse field Ising model(TFIM) and the Kitaev p-wave chain model. Following Kitaev, we will diagonalize the Kiatev chain model using a Majorana fermion representation and show that, in its topological phase, the Kitaev chain model has a Majorana edge model and also topological degeneracy. Majorana fermions are very important in our study, and we will look closely at their algebra and how, as a quantum system, they are different from the standard(Dirac)fermions.\\
The Hamiltonian for the transverse field Ising model is \cite{Sachdev}:
\begin{align}
H=-J\sum_{i=1}^{N-1}\sigma_{i}^{x}\sigma_{i+1}^{x}-h_{z}\sum_{i=1}^{N}\sigma_{i}^{z}
\end{align}
where $J$ is the ferromagnetic exchange constant and $h_{z}$ is the Zeeman field in the $Z$ direction and $\sigma_{i}$ is the Pauli spin matrix at the i-th site. This model has $Z_{2}$ symmetry, due to which the global symmetry operator $\prod_{i}\sigma_{i}^{z}$ commutes with the Hamiltonian. 
\begin{align}
\left[\prod_{i} \sigma_{i}^{z},H\right]=0
\end{align}
The global symmetry operator flips all the spins. There is a doubly degenerate ground  state. This model exhibits two phases that  can be understood on the basis of Landau's symmetry breaking theory. There is a ferromagnetically ordered phase that arises when the symmetry of the model is broken.There is a disordered phase in which symmetry is intact.
We will now apply the Jordan-Wigner transformation to this model to map it to a fermionic model that will turn out to be the Kitaev chain model. The Jordan-Wigner transform maps the spin operators into Fermionic ones:
\begin{align}
c_{i}=\sigma_{i}^{\dag}\left(\prod_{j=1}^{i-1}\sigma_{i}^{z}\right) \quad c_{i}^{\dag}=
\sigma_{i}^{-}\left(\prod_{j=1}^{i-1}\sigma_{i}^{z}\right)
\end{align}
\begin{align}
&H=-t\sum_{i=0}^{N-1}(c_{i}^{\dag}c_{i+1}+h.c.)+\bigtriangleup\sum_{i=0}^{N-1}c_{i}^{\dag}
	c_{i+1}^{\dag}+h.c. \nonumber 
	-\mu\sum_{i=0}^{N}c_{i}^{\dag}c_{i}
\end{align}
where $t,\Delta,\mu$ are hopping strength,superconducting order parameter and chemical potential respectively.
Kitaev employed a Majorana fermion representation to diagonalize this Hamiltonian.
\begin{align}
&c_{i}=\frac{\gamma_{1,i}-i\gamma_{2,i}}{\sqrt{2}} \quad 
c_{i}^{\dag}=\frac{\gamma_{1,i}+i\gamma_{2,i}}{\sqrt{2}}
\end{align}
In the Majorana fermion representation the Hamiltonian gets transformed to:
\begin{align}
&H=it\sum_{i=0}^{N-1}(\gamma_{1,i}\gamma_{2,i+1}-\gamma_{2,i}\gamma_{1,i+1})+\nonumber\\
&i\Delta\sum_{i=0}^{N-1}(\gamma_{1,i}\gamma_{2,i+1}+\gamma_{2,i}\gamma_{1,i+1})
-\mu\sum_{i=0}^{N}(\frac{1}{2}-i\gamma_{1,i}\gamma_{2,i})
\end{align}
The Hamiltonian has trivial phase and topological phase. Trivial phase is obtained for the choice of parameters:  $t=\Delta=0$. In this case two Majorana fermions at each site couple together to form a complex fermion, and there is no topological phase as there are no Majorana edge modes.
Choosing $\mu=0$ and $t=\Delta$ the Hamiltonian becomes.
\begin{align}
H=2it\sum_{i=0}^{N-1}\gamma_{1,i}\gamma_{2,i+1}
\end{align}
We can define a complex fermion:
\begin{align}
a_{i}=\frac{\gamma_{2,i+1}-i\gamma_{1,i}}{\sqrt{2}}
\end{align}
The Hamiltonian becomes:
\begin{align}
H=\left(t\sum_{i=0}^{N-1}a_{i}^{\dag}a_{i}-\frac{1}{2}\right)
\end{align}
We can see that the ground state of this Hamiltonian has no a-fermions. But there is more to the story because there are two Majorana fermions that have not been included in the Hamiltonian. Taking them together we can form another fermion which is  non-local, residing at the two ends of the chain. 
\begin{align}
a_{0}=\frac{\gamma_{1,N}-i\gamma_{2,0}}{\sqrt{2}}
\end{align}
So there is boundary term $H_{b}$ corresponding to two Majorana edge modes. Presence of the boundary term due to the bulk topology is another feature of topological order in which there is a bulk-boundary correspondence.
\begin{align}
H_{b}=\epsilon_{0}a_{0}^{\dag}a_{0} 
\end{align}
where $\epsilon_{0}$ vanishes for the case of infinite chain length. With this boundary term included in the Hamiltonian, we can see that it has a doubly degenerate ground state depending on the presence or absence of the edge mode. It is this edge mode which is the feature of the topological phase of Kitaev chain and is related to the topological invariant of the bulk spectrum. The two ground states can be distinguished by a parity operator. They have even and odd parity respectively. $\mid 0 \rangle$ has no $a_{0}$ fermion and hence an even number of fermions while as $\mid 1\rangle$ has one $a_{0}$ fermion and hence odd parity. So the presence of an edge mode gives rise to double degeneracy. This degeneracy is an example of topological degeneracy because it is protected by a topological invariant. Since the topological invariant comes from a particle-hole symmetry which is a discrete symmetry, such topological order has been called symmetry protected topological order.
\subsection{Majorana fermions versus complex fermions}
Majorana fermions  can be taken algebraically as building blocks of (standard) fermions. The algebra of Majorana fermions makes them very different from the usual fermions. 
\begin{align}
\gamma=\gamma^{\dagger} \quad \gamma^{2}=1 \quad \gamma_{1}\gamma_{2}=-\gamma_{2}\gamma_{1} \quad P=i\gamma_{1}\gamma_{2}
\end{align}
where $\gamma$ is the Majorana fermion operator and P is parity operator for two Majorana fermions. More compactly, we can write:
\begin{align}
\{\gamma_{i},\gamma_{j}\}=2\delta_{ij}
\end{align}
Majoranas are very different from the complex fermions because they are self-hermitian and hence the creation and annihilation operators are the same, which means that a Majorana fermion is its own anti-particle. A fermionic vacuum can not be defined for Majorana fermions because there is no well-defined number operator, or in other words the number of Majorana fermions is not a well-defined quantity, and hence not a quantum number that can be used to label Majorana fermions. Majorana fermions don't have $U(1)$ symmetry, and hence a number operator can not be defined for them. However, they have $Z_{2}$ symmetry; parity is conserved for Majorana fermions.
	
Complex fermions can be mapped to Majorana fermion operators. Let $\gamma_{1}$ and $\gamma_{2}$ be two Majorana fermion operators, then corresponding complex fermion operators are:
\begin{align}
c=\frac{1}{\sqrt{2}}(\gamma_{1}+i\gamma_{2}) \quad c^{\dagger}=\frac{1}{\sqrt{2}}(\gamma_{1}-i\gamma_{2})
\end{align}

When taken abstractly Majorana fermions appear to be very unrealistic particles, but physically they can appear as  Bogoliubov quasiparticles which exist as zero modes in topological superconductors. They are zero energy solutions of the BdG equation and are different from Majorana spinors which are solutions of the Dirac equation.\\
For a general choice of parameters, one finds that the Hamiltonian is an antisymmetric matrix and hence has doubled spectrum. For every energy state there is another degenerate eigenstate.
So one can say that the presence of particle-hole symmetry turns a Hermitian matrix into a real anti-symmetric matrix which has doubled spectrum.
	
The number of Majorana modes is a topological invariant called the $Z_{2}$ invariant, and is given by a Pfaffian of the Hamiltonian. It is also that case that the effects of Majorana fermions in supercooled nanowires has been observed \cite{Kou} in terms of correlations in quantum states of the ends of the wires. A literature has grown around these effects and their possible applications to topological quantum computing.

Fermions obey the Grassmann algebra:
\begin{align}
\{c_{i},c^{\dag}_{i}\}=\delta_{ij} \quad c_{i}^{2}=(c^{\dag}_{i})^{2}=0 \quad N=c^{\dag}c \quad N^{2}=N
\end{align} 
where $c^{\dag}$, $c$ and N are the creation, annihilation and number operators for a fermion.
	\begin{align}
	\mid 1\rangle= c^{\dag}\mid 0\rangle \quad \mid 0\rangle= c\mid 1\rangle \\
	c\mid 0\rangle=c^{\dag}\mid 1\rangle=0 
	\end{align}
	Fermions have a vacuum state. Creation and annihilation operators are used to construct the states of fermions. Fermions have $U(1)$ symmetry, and hence the number of fermions is conserved, and occupation number is a  well-defined quantum number. The number of fermions in a state is given by the eigenvalue of the number operator. Here the number operator is idempotent, and hence there are only two eigenvalues:$0,1$.   Also, different fermion operators anti-commute with each other and hence obey Fermi-Dirac statistics.\\

	\section{Algebra of Majorana doubling}
	In this section we will revisit the algebra of Majorana fermions and see that, in the way it is usually presented, some of the significant higher order products are not used. In the Kitaev paper, the  algebra of Majorana fermions is written as 
	\begin{align}
	\{a_{i},a_{j}\}=2\delta_{ij}
	\end{align}
	This equation defines the Clifford algebra of Majorana fermions. The full algebra is generated by all the ordered products of these operators.   For the case of three Majorana fermions the full Clifford algebra is described below:
	\begin{align}
	&\{ 1,\gamma_{1} = a_1,\gamma_{2}=a_2,\gamma_{3}=a_3,\gamma_{12}=a_1 a_2,\gamma_{23}=a_2 a_3,\nonumber \\
	&\gamma_{31}=a_3 a_1,\gamma_{123}= a_1 a_2 a_3  \}
	\end{align}
	The Clifford algebra of three Majorana fermions is $8$ dimensional, with these eight independent generators. There are three bivectors $\gamma_{12}$,$\gamma_{23}$,$\gamma_{31}$ and one trivector(also called a pseudoscalar) $\gamma_{123}$. Bivectors are related to rotations and trivector will turn out to be very important for our discussion on topological order because it is a chirality operator which distinguishes between even and odd parity. We refer to \cite{Rukhsan} for more discussion on Clifford algebra of spin. In that paper, relation between spin and fermionic systems has been made clear. The reader will find seeds of the duality between Pauli matrices(spin) and Clifford algebra(Majorana fermions).
	
	Lee and Wilczek\cite{Lee} gave an illuminating analysis of the doubled spectrum of the Kitaev chain model. They showed  that the algebra that has been considered for the Kitaev chain model is conceptually incomplete. Using case of three Majorana fermions which are at the edges of superconducting wires, it is shown that the Hamiltonian of these Majorana fermions has more algebraic structure than anticipated. The difference lies in another Majorana operator which  has been called an \textit{Emergent Majorana} for the reason that it obeys all the properties of a Majorana fermion. We briefly review their analysis here and later on generalize it.
	
	Let $b_{1}$,$b_{2}$ and $b_{3}$ be three Majorana fermions that can occur at the ends of three p-wave superconducting nano-wires. This situation is not artificial,  rather it is crucial for the topological Kondo effect\cite{Beri} where  three Majorana fermions give rise to non-local spin-$1/2$ object which then couples to the conduction fermions in the leads. Similarly, three such wire junctions have also been explored to study the Kitaev spin model\cite{Kells2,Yao}. 
	\begin{align}
	\{b_{j},b_{k}\}=2\delta_{jk}
	\label{b_Majorana}
	\end{align}
	We can write down a Hamiltonian for these interacting Majoranas coming from three different wires.
	\begin{align}
	H_{m}=-i(\alpha b_{1}b_{2}+\beta b_{2}b_{3} + \gamma b_{3}b_{1})
	\label{Three-eta-H}
	\end{align}
	Now it is known that Majorana bilinears generate a spin algebra so one would naively think that it is a spin Hamiltonian. But the spin Hamiltonian neither has edge modes nor any topological degeneracy. To understand this, one needs to realize that the Clifford algebra generated by Majorana fermions is larger than what is present in equation~\ref{b_Majorana}. There are other generators of the algebra. The Hamiltonian given in equation~\ref{Three-eta-H} has parity symmetry due to which fermion number $N_{e}$ is conserved modulo 2. 
	\begin{equation}
	[H_{m},P]=0 \quad P= (-1)^{N_{e}} \quad P^{2}=1
	\end{equation}
	Physically the full implications of the parity operator need to be taken into consideration to conceptually complete the algebra. There is a special operator $\Gamma$ in the algebra which we call as \textit{Emergent Majorana} because it has all the properties of a Majorana fermion.
	\begin{align}
	&\Gamma \equiv -ib_{1}b_{2}b_{3}\\
	&\Gamma^{2}=1 \quad [\Gamma,b_{j}]=0 \quad [\Gamma,H_{m}]=0 \quad \{\Gamma,P\}=0
	\end{align}
	The emergent Majorana operator commutes with the Hamiltonian, and hence there is an additional symmetry present, as it anti-commutes with the parity operator and hence it shifts among the parity states. Both the $P$ and $\Gamma$ operators commute with Hamiltonian but anti-commute with each other due to which there is doubling of the spectrum. The presence of the this extra symmetry leads to the doubled spectrum. This doubling is different from Kramer's doubling\cite{Bernevig} because no time reversal symmetry is needed.
	In the basis in which $P$ is diagonal with $\pm 1$ eigenvalues, the $\Gamma$ operator takes the states into degenerate eigenstates with eigenvalues $\mp 1$.
	It is very suggestive to write the parity operator for $H_{m}$ in terms of a $\Gamma$ operator in line with\cite{Akhmerov,Feldman}. The case of an odd number of Majorana fermions can always be thought as a combination of complex fermions($c$ fermions) and one Majorana fermion so that the parity operator can be written as $P=\prod_{i}(1-2c_{i}^{\dag}c_{i})\gamma$. This parity operator constitutes the local parity operator of reference \cite{Akhmerov}.
	
	Two Majorana fermions combine together to give a Dirac fermion. Three Majorana fermions is the simplest case that shows non-trivial Majorana physics. So equation~\ref{Three-eta-H} gives the simplest Hamiltonian that we can write down for Majorana fermions having non-trivial spectral and hence physical properties. We have seen for this Hamiltonian that there are additional symmetries which lead not only to the topological degeneracy for the ground state, but the Hamiltonian has doubling for the whole spectrum including excited states. This is a very powerful implication of the existence of emergent Majorana fermions and hence of the emergent symmetry operators. We can not talk about topological order for three Majorana fermions, but later on we will see that it is the presence of these emergent fermionic symmetries which leads to the topological order in Majorana fermion models.
	
	Emergent Majorana fermions were considered in \cite{Akhmerov} and it was shown that they are more robust and hence better candidates for topological quantum computing. In \cite{Feldman} authors have also studied the Hamiltonian given in equation~\ref{Three-eta-H} and using the anti-symmetry property of the Hamiltonian arrived at what we call as emergent Majorana fermions.

	The doubled spectrum of the Kitaev chain Hamiltonian comes from this algebraic structure which leads to extra symmetries. This algebraic structure is non-perturbative, and hence is robust to perturbations as long as they preserve the discrete symmetry.
	\subsection{Emergent Supersymmetry in Majorana Fermions models}
	In this section we will show that how the $\Gamma$ operator leads to supersymmetry(SUSY). Supersymmetry has been studied in lattice models for fermions\cite{Fendley}(and references therein). Recently, there has been some interest in searching for the supersymmetry in lattice models for Majorana fermions\cite{Armin,Grover,Zhang2}. However, in those situations SUSY arises only at the critical point of the model. In our case, we find emergent supersymmetry which is tied to topological order because we need $\Gamma$ operator which exists only in topological phase. This SUSY was already pointed out by Lee and Wilczek. Emergent supersymmetry  was  also found in \cite{Hseih} for Majorana fermion models with translational symmetry. The difference with our case is that we need to have a Majorana mode operator $\Gamma$ rather than translational symmetry to have emergent SUSY. It needs to be noted that the doubling of the spectrum is not due to the supersymmetry.
	
	We once again consider a system of an odd number of Majorana fermions so that we can define the $\Gamma$ operator. For the quadratic Hamiltonian for Majorana fermions, parity is conserved and hence we can define parity operator, P which commutes with Hamiltonian. We define $N=2$ supersymmetry generator, Q which commutes with Hamiltonian and hence shows that Majorana fermion Hamiltonian has supersymmetry.
	\begin{align}
	&Q=\sqrt{H}\left(\frac{1+P}{2}\right)\Gamma \\
	&Q^{2}=0 \quad \left[Q,H\right]=0 \quad \left \{Q,Q^{\dag}\right\}=H
	\end{align}
	It needs to be noted that we did not need to have translation symmetry to have supersymmetry in our Hamiltonian. So in that sense we show that supersymmetry  is a feature of Majorana fermion Hamiltonians and is also tied to their topological order because we need the existence of the $\Gamma$ operator. 
	\subsection{Majorana Zero Modes and $\Gamma$ operator}
	What we have already found is that the complete algebra of Majorana fermions have extra operators that have been called  emergent Majorana fermions and represented by $\Gamma$ operators.  For the case of three Majorana fermions we find that the $\Gamma$ operator is the 
	emergent Majorana fermion and is also the symmetry of the Hamiltonian. In this section we will see that the $\Gamma$ operator is also the fermionic zero mode of the Hamiltonian. Fermionic zero modes give a clear signature of topological order\cite{Fendley}. First, we give a definition of the fermionic zero mode, and then we will show how that is related to Emergent Majorana fermions or the $\Gamma$ operator. Following\cite{Fendley}, a fermionic zero mode is an operator $\Gamma$ such that $\Gamma$
	\begin{itemize}
		\item Commutes with Hamiltonian: $[H,\Gamma]=0$
		\item anticommutes with parity: $\{P,\Gamma\}=0$
		\item has finite "normalization" even in the $L \rightarrow \infty$ 
		limit:  $\Gamma^{\dag}\Gamma=1$.
	\end{itemize}
	Now we can easily see that the first two properties are the defining properties of the $\Gamma$ operator and hence are satisfied. $\Gamma$, like a Majorana operator, squares to unity and so is always normalized. 
	So our $\Gamma$ operator satisfies all the properties of the zero edge mode. This Majorana zero mode has also been discussed in\cite{Akhmerov,Feldman}. In \cite{Akhmerov} it was shown that these Majorana zero modes are more robust to the environment and hence should be used for quantum computing. $\Gamma$ operator is the local parity operator corresponding to Majorana zero mode. In \cite{Feldman} the authors derived the existence of this Majorana fermion mode by looking at the symmetry property of the Hamiltonian for the three Majorana fermions. As we can see, that Hamiltonian is an anti-symmetric matrix and has zero eigenvalue.

	In the next section we will consider a general case of an odd number of Majorana fermions and show that there are Majorana mode operators and hence the topological order in the corresponding Hamiltonian.
	\section{Topological Order and Fermionic Mode Operators}
	We  consider general case of odd number $N$ $>$3 of Majorana fermions. We will find that, as for the case of three Majorana fermions, there are emergent Majorana fermions and their Majorana mode operators are the symmetries of the Hamiltonian. For the case of three Majorana fermions there was only one emergent Majorana mode operator. However, for the general case there are many more Majorana mode operators as shown in\cite{Chamon}. We will focus just on the generalized case of the $\Gamma$ operator considered in previous section.
	
	We consider a system that has $2N+1$ Majorana fermions. These Majorana fermions will span a vector space of dimensionality $2^{2N+1}$ corresponding to the number of linearly independent generators of the Clifford algebra\cite{Lounesto,Snygg}. These generators can be written as
	\begin{align}
	1,\gamma_{1},\gamma_{2}...,\gamma_{2N+1},\nonumber \\
	\gamma_{1}\gamma_{2},\gamma_{1}\gamma_{3}....\\
	\gamma_{1}\gamma_{2}\gamma_{3}....\\
	\vdots \\
	\gamma_{1}\gamma_{2}.....\gamma_{2N+1}
	\label{Clifford}
	\end{align}
	In addition to the Majorana fermion operators, all the higher order products are also the generators of the Clifford algebra corresponding to $2N+1$ Majorana fermions. It is these higher order generators of the Clifford algebra that also play a very important role in the dynamics of the Majorana fermion systems. Based on these generators of the Clifford algebra of Majorana fermions, we can write down the Hermitian operators which will commute with the Hamiltonian and hence are the symmetries of the Hamiltonian. These symmetries are the extra symmetries in addition to the $Z_{2}$ symmetry present in any quadratic Majorana fermion Hamiltonian. These symmetries are the emergent symmetries generated by higher order Majorana fermion generators.
	
	The most general local quadratic Hamiltonian for the Majorana fermions can be written as
	\begin{align}
	H=i\sum_{ij}h_{ij}\gamma_{i}\gamma_{j}
	\label{Quad-H}
	\end{align}
	Due to the anti-commuting nature of the Majorana fermions, $h_{ij}=-h_{ji}$, this Hamiltonian has manifest $Z_{2}$ symmetry and consequently the Hamiltonian can be diagonalized in the parity eigenbasis. Since the Hamiltonian is bilinear in Majorana fermion operators, all the elements of the algebra(of equation \ref{Clifford}) with even parity will commute with Hamiltonian and hence are the emergent symmetries of it. To have topological order, there must exist Majorana mode operators as defined in section $3.2$. As we defined the Majorana mode operator for the case of three Majorana fermions, similarly we find that there is corresponding emergent Majorana fermion operator as written below:
	\begin{align}
	\Gamma=i^{(2N+1)2N}\gamma_{1}\gamma_{2}...\gamma_{2N+1} \quad \Gamma^{2}=1 \quad [\Gamma,H]=0
	\end{align}
	Being the product of all Majorana fermions, it commutes with the Hamiltonian. It also squares to unity and anti-commutes with the parity operator. Hence it satisfies all the properties of a Majorana mode operator, and hence we show that the quadratic Majorana fermion model exhibits topological order. There are other Majorana mode operators in addition to the ones that we have written down above as shown in \cite{Chamon,Feldman}.
	
	The quadratic Majorana fermion model as given in equation \ref{Quad-H} has manifest $Z_{2}$ symmetry but there are additional symmetries as shown above. We call these symmetries  \textit{emergent symmetries} while the $Z_{2}$ symmetry is the microscopic symmetry of the Hamiltonian, and it leads to the parity(fermion number) conservation. In the parity basis, the Hamiltonian takes quite nice form as given\cite{Kells}.
	\begin{align}
	H=\epsilon(\mid e\rangle\langle e\mid +\mid o\rangle\langle o\mid)
	\end{align}
	where $\mid e\rangle$ ,$\mid o\rangle$ are even and odd parity states. In this notation, Majorana mode operators take conceptually intuitive form:
	\begin{align}
	\Gamma=(\mid e\rangle\langle o\mid +\mid o\rangle\langle e\mid)
	\end{align}
	This form of $\Gamma$ brings out its property of flipping the parity state because it anti-commutes with the parity operator. So the matrix structure of the Majorana fermion model is not just block diagonal with even and odd parity blocks, rather it has two more blocks corresponding to the identity operator and $\Gamma$ operator, both of which occupy a single element block in the Hamiltonian matrix.
	
	Based on our analysis of the quadratic Majorana model, we can now understand why there is topological order in the Kitaev chain model. Note first that $Z_{2}$ symmetry is present in both the Kitaev chain model  and its Jordan-Wigner dual spin model. We have also seen that for the special case of parameters of the Kitaev chain model, it reduces to a quadratic Hamiltonian of the kind that we have studied above, and hence the same analysis holds true for it also. We can now see that the topological phase arises in the Kitaev chain model because there are Majorana mode operators that commute with the Hamiltonian and anti-commute with the parity operator. For the spin model, there are no such symmetries which can lead to topological order.
	So we find that topological order in the Kitaev chain model  arises due to  the enrichment of the emergent symmetries generated by higher order products of the Majorana fermion operators. What we have found goes beyond the duality between Landau order in TFIM and the Kitaev chain model. Based on duality there is a transformation of local observables to non-local observables but we can not find the the generators of the emergent symmetries that lead to topological order in Kitaev chain model.
	
	
	In another words, in the case of the Majorana fermion Hamiltnian there are many conserved quantities whose Liouvillian vanishes and they are conserved under Liouvillian dynamics. This picture can be compared to the one presented in \cite{Fendley3} where the protection of the Majorana edge modes has been shown to be related to prethermalization.
	
	\subsection{Interactions}
	Now we consider  the Hamiltonian with quartic fermion terms and show that emergent Majorana operators exist in the presence of interactions as well.
	\begin{align}
	H=i\sum_{ij}h_{ij}\gamma_{i}\gamma_{j}+\sum_{ijkl}V_{ijkl}\gamma_{i}\gamma_{j}\gamma_{k}
	\gamma_{l}
	\end{align}
	The first term of the Hamiltonian is the same as equation \ref{Quad-H}. The second term is the interaction term where $V_{ijkl}$ is real and anti-symmetric under odd permutations. A first thing to be noted is that the interaction term does not break parity symmetry and hence we can still decompose the Hamiltonian into even and odd parity blocks. However, due to the interaction effects, Majorana mode operators are no longer the same as the ones for the mean-field Hamiltonian, as calculated in section $2$ above. There will be dressing of Majorana mode operators due to the interactions. This is because the Majorana mode operators have to commute with the Hamiltonian and hence the Majorana mode operator will be a linear combination of products of an odd number of Majorana fermions.
	
	The generalized Majorana mode operator  is a dressed operator which gets a contribution from higher order Majorana fermion operators.
	\begin{align}
	O=\sum_{i}u_{i}\gamma_{i}+\sum_{ijk}u_{ijk}\gamma_{i}\gamma_{j}\gamma_{k}+...
	\end{align}
	The $\Gamma$ operator is the product of all Majorana fermions and hence commutes with the Hamiltonian. This Majorana mode operator has a many-body character and survives the effects of interactions as was confirmed in numerical studies also\cite{Kells2}. The reason for that is, the $\Gamma$ operator arises due to the non-perturbative effects of algebra of Majorana fermions as elucidated in section $3$ on the algebra of Majorana doubling.
	\subsection{Disorder}
	The effect of disorder on topological order is another important question that needs to be addressed in order to understand topological order in realistic systems. Although we have not explored the effect of disorder on topological order, we have made some observations that we discuss in this section. We find that there is a striking similarity between topological order and many-body localized phase(MBL) in the sense that MBL is characterized by proliferation of local conserved quantities\cite{Louk,Imbrie}. We have already seen in section $4$ that in case of Majorana fermions models, topological ordered phase has many emergent symmetries and hence conserved quantities.
	This formal similarity between topologically ordered systems and MBL phase in disorderd systems is quite suggestive. In that regard we find recent work\cite{Fendley3} that helps to understand this proliferation of conserved quantities both in the topologically ordered and the MBL phase. In \cite{Fendley3} it was shown that the existence of the Majorana mode operators is a non-equilibrium dynamical phenomenon and is related to prethermalization. We speculate that the emergent symmetries which we have found for Majorana fermion models, may be the reason for pre-thermalization. 
	\section{Symmetry algebra of Topological Protection}
	In an earlier section we have shown that in Majorana Fermion models there are emergent symmetries and especially the symmetries associated with fermionic mode operators. In this section we will take an even more general approach to topological protection based on two sets of symmetry operators. Our approach is related to recent works by Finkelstein et al(citation needed).
	Let ${P}$ and ${Q}$ be two sets of operators such that
	\begin{align}
	[P,H]=[Q,H]=0 \quad \lbrace P,Q\rbrace =0 \quad P^{2}=Q^{2}=0
	\end{align}
	P and Q are symmetry operators of the Hamiltonian H that anti-commute with each other.
	Because P and Q commute with H, so they will have the same eigenstates,  but because P and Q anti-commute, the eigenvalues can not be same.
	\begin{align}
	P\mid\Psi\rangle=m\mid\Psi\rangle  \quad Q(P\mid\Psi\rangle)=mQ(\mid\Psi\rangle) \quad P(Q\mid\Psi\rangle)=-m(Q\mid\Psi\rangle)
	\end{align}
	
	For every state with eigenvalue m, there is another state with eigenvalue -m and hence there will be a doubling of the spectrum. This is the doubling that we found for the Majorana Hamiltonian in section 3. For that case, the two operators are the Parity operator and the $\Gamma$  operator. 
	
	\begin{align}
	P=i\Gamma f \quad Q=\Gamma \quad [P,Q]=[Q,H]=0 \quad \{P,Q\}=0 \quad P^{2}=Q^{2}=0
	\end{align}
	This degeneracy is different than the one which we have for a quantum system whose Hamiltonian commutes with various symmetry operators. Those degeneracies are susceptible to local perturbation that leads to the lifting of the degeneracy. This degeneracy is protected by non-local symmetry operators and these symmetries can not be broken by local perturbations. The algebraic structure of equation 27 ensures the topological protection of the ground state and the encoded information. As long as, these symmetries as represented by P and Q operators are present, quantum information stored in the degenerate ground state is also preserved. It is very interesting to see that similar algebraic structure has been found in \cite{Ioffe}. In that work, the the same question of the topological protection of quantum information and the necessary algebraic structure has been explored. Our main interest has been Majorana fermion Hamiltonians,  whereas they have considered a lattice spin model. They have shown that, even for a lattice spin model, the same algebraic structure of symmetry operators ensures the topological protection of the ground state -  which can be used as topological qubit. One obvious difference is that they use more operators because they have a two dimensional lattice model that has more degrees of freedom and a larger Hilbert space.

	\section{Topological order and Yang-Baxter equation}
	Majorana fermions have been the focus of interest in research in topological quantum computation because as shown in \cite{Moore,Ivanov} that Majorana fermions have non-abelian braid statistics and generate representation of braid group. Kitaev chain realization of Majorana fermions have given ways to engineer Majorana fermions and there has already been some progress on that front\cite{Mourik}.
	It has also been realized\cite{Ge} that the Majorana representation of braid group is different than the ones known in the literature. This representation has been called a type-II representation. Now the question which has been asked is that is the topological order which arises from quantum entanglement also related to topological entanglement which arises from the solutions of Yang-Baxter equation. Majorana fermions give new solutions to Yang-Baxter equations and hence the new type of topological entanglement. When there is topological order, we get a representation of braid group and also solutions to YBE. We will first briefly review the Majorana fermion representation of braid group\cite{Ivanov,Kauffman}. 
	. Then we will discuss about the new solution of Yang-Baxter equation(YBE)which has been called type-II solution. One important property of type-II solution is its R matrix commutes with the $\Gamma$ operator which is Majorana edge mode operator.
	
	Braiding operators arise from a row of Majorana Fermions $\{ \gamma_{1}, \cdots \gamma_{n} \}$ as follows: Let 
	\begin{align}
	\sigma_{i} = (1/\sqrt{2})(1 + \gamma_{i+1}\gamma_{i})
	\end{align}
	Note that if we define 
	\begin{align}
	\lambda_{k} = \gamma_{i+1} \gamma_{i}
	\end{align}
	for $i = 1,\cdots n$ with $\gamma_{n+1} = \gamma_{1}$, then 
	\begin{align}\lambda_{i}^2 = -1
	\end{align}
	and
	\begin{align}
	\lambda_{i} \lambda_{j} +  \lambda_{j} \lambda_{i} = 0
	\end{align}
	where $i \ne j.$ From this it is easy to see that 
	\begin{align}
	\sigma_{i}\sigma_{i+1}\sigma_{i} = \sigma_{i+1}\sigma_{i}\sigma_{i+1}
	\end{align}
	for all $i$ and that 
	\begin{align}
	\sigma_{i} \sigma_{j} = \sigma_{j}\sigma_{i}
	\end{align}
	when $|i-j|>2.$ Thus we have constructed a representation of the Artin braid group from a row of Majorana fermions. This construction is due to Ivanov \cite{Ivanov} and he notes that 
	\begin{align}
	\sigma_{i} = e^{(\pi/4) \gamma_{i+1} \gamma_{i} }
	\end{align}
	
	In \cite{Ge} authors make the further observation that if we define 
	\begin{align}
	\breve{R}_{i}(\theta) = e^{\theta \gamma_{i+1} \gamma_{i} }
	\end{align}
	Then $\breve{R}_{i}(\theta)$ satisfies the 
	full Yang-Baxter equation with rapidity parameter $\theta.$ That is, we have the equation 
	\begin{align}
	\breve{R}_{i}(\theta_{1})\breve{R}_{i+1}(\theta_{2})\breve{R}_{i}(\theta_{3}) = \breve{R}_{i+1}(\theta_{3})\breve{R}_{i}(\theta_{2})\breve{R}_{i+1}(\theta_{1})
	\end{align}
	This makes if very clear that $\breve{R}_{i}(\theta)$ has physical significance, and suggests examining the physical process for a temporal evolution of the unitary operator  $\breve{R}_{i}(\theta).$\\
	In fact, following \cite{Ge},  we can construct a Kitaev chain based on the solution $\breve{R}_i(\theta)$ of the Yang-Baxter Equation. Let a unitary evolution be governed by $\breve{R}_i(\theta)$. When $\theta$ in the unitary operator 
	$\breve{R}_i(\theta)$ is time-dependent, we define a  state $|\psi(t)\rangle$ by $|\psi(t)\rangle=\breve{R}_i|\psi(0)\rangle$. With the Schr\"{o}dinger equation $i\hbar\tfrac{\partial}{\partial t}|\psi(t)\rangle=\hat{H}(t)|\psi(t)\rangle$ one obtains:
	\begin{equation}
	i\hbar\tfrac{\partial}{\partial t}[\breve{R}_i|\psi(0)\rangle]=\hat{H}(t)\breve{R}_i|\psi(0)\rangle.
	\end{equation}
	Then the Hamiltonian $\hat{H}_i(t)$ related to the unitary operator $\breve{R}_i(\theta)$ is obtained by the formula:
	\begin{equation}\label{SchrodingerEquation}
	\hat{H}_i(t)=\textrm{i}\hbar\tfrac{\partial\breve{R}_i}{\partial t}\breve{R}_{i}^{-1}.
	\end{equation}
	Substituting $\breve{R}_i(\theta)=\exp(\theta\gamma_{i+1}\gamma_{i})$ into equation (\ref{SchrodingerEquation}), we have:
	\begin{equation}\label{2MFHamiltonian}
	\hat{H}_i(t)=\textrm{i}\hbar\dot{\theta}\gamma_{i+1}\gamma_{i}.
	\end{equation}
	This Hamiltonian describes the interaction between $i$-th and $(i+1)$-th sites via the parameter $\dot{\theta}$. When $\theta=n \times \tfrac{\pi}{4}$, the unitary evolution corresponds to the braiding progress of two nearest Majorana fermion sites in the system as we have described it above. Here $n$ is an integer and signifies  the time of the braiding operation. We remark that it is interesting to examine this periodicity of the appearance of the 
	topological phase in the time evolution of this Hamiltonian. For applications, one may consider processes that let the Hamiltonian take the the system right to one of these topological points and then this Hamiltonian cuts off.
	One may also think of a mode of observation that is tuned in frequency with the appearances of the topological phase.\\  
	
	In \cite{Ge} authors also point out that if we only consider the nearest-neighbour interactions between Majorana Fermions,  and extend equation (\ref{2MFHamiltonian}) to an inhomogeneous chain with $2N$ sites, the derived model is expressed as:
	\begin{equation}\label{YBEKitaev}
	\hat{H}=\textrm{i}\hbar\sum_{k=1}^{N}(\dot{\theta}_1\gamma_{2k}\gamma_{2k-1}+\dot{\theta}_2\gamma_{2k+1}\gamma_{2k}),
	\end{equation}
	with $\dot{\theta}_1$  and $\dot{\theta}_2$ describing odd-even and even-odd pairs, respectively.

	They then analyze the above chain model in two cases:
	
	\begin{enumerate}
		\item $\dot{\theta}_1>0$, $\dot{\theta}_2=0.$
		
		In this case, the Hamiltonian is:
		\begin{equation}\label{YBEtrivil}
		\hat{H}_1=\textrm{i}\hbar\sum_{k}^{N}\dot{\theta}_1\gamma_{2k}\gamma_{2k-1}.
		\end{equation}
		The Majorana operators $\gamma_{2k-1}$ and $\gamma_{2k}$ come from the same ordinary fermion site k, $\textrm{i}\gamma_{2k}\gamma_{2k-1}=2a_{k}^{\dag}a_{k}-1$ ($a_{k}^{\dag}$ and $a_{k}$ are spinless ordinary fermion operators). $\hat{H}_1$ simply means the total occupancy of ordinary fermions in the chain and has U(1) symmetry, $a_j\rightarrow e^{i\phi}a_j $.  Specifically, when $\theta_1(t)=\tfrac{\pi}{4}$, the unitary evolution $e^{\theta_{1}\gamma_{2k}\gamma_{2k-1}}$ corresponds to the braiding operation of two Majorana sites from the same k-th ordinary fermion site.  The ground state represents the ordinary fermion occupation number 0. In comparison to 1D Kitaev model, this Hamiltonian corresponds to the trivial case of Kitaev's. This Hamiltonian is described by the intersecting lines above the dashed line, where the intersecting lines correspond to interactions. The unitary evolution of the system $e^{-i{\int\hat{H}_1dt}}$ stands for the exchange process of odd-even Majorana sites.\\
		
		\item $\dot{\theta}_1=0$, $\dot{\theta}_2>0.$
		
		In this case, the Hamiltonian is:
		\begin{equation}\label{YBEtopo}
		\hat{H}_2=\textrm{i}\hbar\sum_{k}^{N}\dot{\theta}_2\gamma_{2k+1}\gamma_{2k}.
		\end{equation}
		This Hamiltonian corresponds to the topological phase of 1D Kitaev model and has $\mathbb{Z}_2$ symmetry, $a_j\rightarrow -a_j$. Here the operators $\gamma_1$ and $\gamma_{2N}$ are absent in $\hat{H}_2.$ The Hamiltonian has two degenerate ground state, $|0\rangle$ and $|1\rangle=d^{\dag}|0\rangle$, $d^{\dag}=e^{-i\varphi/2}(\gamma_1-i\gamma_{2N})/2$. This mode is the so-called Majorana mode in 1D Kitaev chain model. When $\theta_2(t)=\tfrac{\pi}{4}$, the unitary evolution $e^{\theta_{2}\gamma_{2k+1}\gamma_{2k}}$ corresponds to the braiding operation of two Majorana sites $\gamma_{2k}$ and $\gamma_{2k+1}$ from $k$-th and $(k+1)$-th ordinary fermion sites, respectively.
		
		\medskip
	\end{enumerate}
	
	Thus the Hamiltonian derived from $\breve{R}_{i}(\theta(t))$ corresponding to the braiding of nearest Majorana fermion sites is exactly the same as the $1D$ wire proposed by Kitaev, and $\dot{\theta}_1=\dot{\theta}_2$ corresponds to the phase transition point in the ``superconducting'' chain. By choosing different time-dependent parameter $\theta_1$ and $\theta_2$, one finds that the Hamiltonian $\hat{H}$ corresponds to different phases.
	These observations of Mo-Lin Ge give physical substance and significance to the Majorana Fermion braiding operators discovered by Ivanov \cite{Ivanov}, putting them into a robust context of Hamiltonian evolution via the simple Yang-Baxterization  $\breve{R}_{i}(\theta) = e^{\theta \gamma_{i+1} \gamma_{i} }.$  Yu and Mo-lin Ge\cite{Ge} make another observation, that we wish to point out. In \cite{Kauffman}, Kauffman and Lomonaco observe that the Bell Basis Change Matrix in the quantum information context is a solution to the Yang-Baxter equation. Remarkably this solution can be seen as a $4 \times 4$ matrix representation for the operator $\breve{R}_{i}(\theta).$\\ 
	
	This lets one can ask whether there is relation between topological order and quantum entanglement and braiding \cite{Kauffman} which is the case for the Kitaev chain where non-local Majorana modes are entangled and also braiding.\\
	
	The Bell-Basis Matrix $B_{II}$ is given as follows:
	\begin{equation}
	B_{II}=\frac{1}{\sqrt{2}}\left[\begin{array}{cccc}
	1 & 0 & 0 & 1\\
	0 & 1 & 1 & 0\\
	0 & -1 & 1 & 0\\
	-1 & 0 & 0 & 1
	\end{array}\right]=\frac{1}{\sqrt{2}}\bigl(I+M\bigr)\quad\bigl(M^{2}=-1\bigr)
	\end{equation}
	and
	\begin{eqnarray}
	M_{i}M_{i\pm1}&=&-M_{i\pm1}M_{i},\quad M^{2}=-I,\\
	M_{i}M_{j}&=&M_{j}M_{i,}\quad \big|i-j\big|\geq2.
	\end{eqnarray}

	\noindent{\bf Remark.}  The operators $M_{i}$ take the place here of the products of Majorana Fermions $\gamma_{i+1}\gamma_{i}$ in the Ivanov picture of braid group representation in the form
	$$\sigma_{i} = (1/\sqrt{2})(1 + \gamma_{i+1}\gamma_{i}).$$ This observation of authors in \cite{Ge}  gives a concrete interpretation of these braiding operators and relates them to a Hamiltonian for the physical system.
	This goes beyond the work of Ivanov, who examines the representation on Majoranas obtained by conjugating by these operators. The Ivanov representation is of order two, while this representation is of order eight.
	The reader may wish to compare this remark with the contents of \cite{Kauffman1} where we associate Majorana fermions with elementary periodic processes. These processes can be regarded as prior to the 
	periodic process associated with the Hamiltonian of Yu and  Mo-Lin Ge\cite{Ge}.
	
	\noindent{\bf Remark.} We write down a Majorana fermion representation of Temperley-Lieb algebra(TLA) which is related to the Braid group representation discussed above. We define $A$ and $B$ as $A=\gamma_{i}\gamma_{i+1}$, $B=\gamma_{i-1}\gamma_{i}$ where $A^{2}=B^{2}=-1$. Note the following relations:
	\begin{align}
	&U=(1+iA) \quad V=(1+iB),\\
	& U^{2}=2U \quad V^{2}=2V,\\
	&UVU=V \quad VUV=U,
	\end{align}
	Thus a Majorana fermion representation of TLA is given by:
	\begin{align}
	&U_{k}=\frac{1}{\sqrt{2}}(1+i\gamma_{k+1}\gamma_{k}),\\
	&U_{k}^{2}=\sqrt{2}U_{k},\\
	&U_{k}U_{k\pm 1}U_{k} = U_{k},\\
	&U_{k}U_{j}= U_{j}U_{k} \quad for |k-j|\ge 2.
	\end{align}
	Hence we have a representation of the Temperley-Lieb algebra with loop value $\sqrt{2}.$ Using this representation of the Temperley-Lieb algebra\cite{Kauffman2,Kauffman3}, we can construct (via the Jones representation of the braid group to the Temperley-Lieb agebra) another represention of the braid group that is based on Majorana Fermions. It remains to be seen what is the physical significance of this new representation.
	\subsection{Topological order and topological entanglement}
	One of the main aims of this paper is to understand the relation between the topological order which comes from quantum entanglement and topological entanglement which comes from braid group representation and Yang-Baxter equation. This is an extension of the work\cite{Kauffman,Kauffman1} in which it was shown how quantum entanglement is related to braid group and Yang-Baxter equation. So it became natural to understand topological order based on the approach of \cite{Kauffman,Kauffman1,Kauffman2} . In this section, we will show that Yang-Baxter equation based approach opens new ways to understand the topological order in case of Majorana fermion models. However, this approach is not restricted to Majorana fermions only rather it is a very general approach which is not limited to quadratic Hamiltonians and hence offers a new method for the classification of the topological phases, which goes beyond K theory and Berry phase based methods.
	
	To understand the relation between quantum entanglement in the Kitaev chain model and the corresponding topological entanglement which manifests as braid group representation, we point out that \textit{ it is only in the topological phase of the Kitaev chain model that braid group representation arises while as in topologically trivial phase there are no Majorana edge modes and hence no braid group representation.}
	To see this relation mathematically, we rewrite the Kitaev chain Hamiltonian corresponding to topological phase.
	\begin{align}
	H=2it\sum_{i=0}^{N-1}\gamma_{1,i+1}\gamma_{2,i}
	\end{align}
	and now find out that for Majorana representation as shown by Ivanov we need the operator of the form $\frac{1+\gamma_{i+1}\gamma_{i}}{\sqrt{2}}$ which arises only in topological phase. So this brings out the relation between topological order and the topological entanglement(braiding). The solution of the Yang-Baxter equation which arises in the topologically ordered phases of the Majorana fermion models is different than the ones which arise from the other Hamiltonians which do not exhibit topological order. Majorana fermion solutions are called type-II while as the other ones are called type-I solutions\cite{Ge}. Hence the Majorana fermion braiding solutions of 
	Yang-Baxter equation characterize and classify the topologically ordered phases.
	Using this relation we give a new characterization of topological order. \textit{A system is said to be topologically ordered if it gives a type-II solution to Yang-Baxter equation}. This characterization is very general and just depends on the braiding properties of anyons(Majorana fermions in this case) and hence should apply to other systems as well.
	\section{Summary}
	In this paper, we have explored the topological order in Majorana fermion models in one dimension in the context of the Kitaev p-wave chain model that exhibits $Z_{2}$ topological order,  while its spin model equivalent has only Landau order. We find that to have topological order in these Majorana fermion models, there should be fermionic symmetries whose generators are Majorana mode operators.  In addition to these symmetries, the Hamiltonian for Majorana fermions has many emergent symmetries. The emergence of these symmetries in topological order is quite in contrast to the symmetry broken order which involves the breaking of the symmetry of the Hamiltonian. We have found that supersymmetry also emerges in the topological phase of Majorana fermion model. This supersymmetry is different from other supersymmetries found in the Majorana fermion models because the latter ones either emerge at the critical point of the model or need other symmetries such as  time reversal or translational symmetry. The reason why there are more symmetries on the fermionic side of the duality between Kitaev chain model and its Jordan-Wigner dual spin model is that the algebra of Majorana fermions is larger than the spin algebra of the spin model. The Clifford algebra of Majorana fermions has higher order products that commute with the Hamiltonian and hence are the emergent symmetries of the model and lead to the block diagonal matrix structure of the Majorana fermion model. There is a special element in the algebra which is the product of all Majorana fermions and has odd parity. We call it the $\Gamma$ operator, and it is the Majorana mode operator and has a many-body character due to its non-local nature. Its existence is the clear signature of the topological order in the system. The $\Gamma$ operator is  also crucial for having supersymmetry that is tied to the topological order.
	
	We have also explored the effect of interactions on the Majorana mode operators and hence on the topological order. We find that in the presence of the interactions, the Majorana mode operators become dressed and become the linear combination of all Majorana fermion operators with odd parity. In the presence of the interactions, the Majorana fermion Hamiltonian has a four block structure, two of them being even and odd parity blocks. The other single element blocks are those of the identity and the $\Gamma$ operator, which both commute with Hamiltonian. We also notice the similarity between the proliferation of the conserved quantities in the topological order in Majorana fermion models and Many-body localized phase(MBL) which is characterized by the existence of the local integrals of motion. This similarity is suggestive of some similar phenomenon leading to these seemingly unrelated phenomena. In that regard, we find recent work of Fendley and collaborators quite insightful. They have shown that existence of Majorana mode operators and hence the topological order in these models is a non-equilibrium dynamical phenomenon and is related to pre-thermalization.
	
	We also explored the implications of emergent symmetries of Majorana fermion models on the braid group representation. We find that braid group generators commute with the Majorana mode operator and hence give a new solution to Yang-Baxter equation, which has been called type-II solution. We also write down a Majorana Fermion presentation of Temperley-Lieb algebra. The fact that a Majorana fermion representation of braid group is different from the type-I solution, which is for spin models, shows that we can distinguish between topological order and Landau order based on the solutions of Yang-Baxter equation. This gives a nice mathematical procedure to check for the topological order. For the Majorana fermion models including the Kitaev chain model, we notice that solutions to the Yang-Baxter equation exist only in the topological phase. So the Yang-Baxter equation can be used to explore topological order in the quantum Hamiltonians. The relation between the Yang-Baxter equation and topological order shows that topological order that arises due to the quantum entanglement is related to topological entanglement of the braid group.
	
	Understanding topological order is very important not only for topological quantum computation, but is also very important within condensed matter physics where more and more systems are being discovered that exhibit topological order. Our work is very significant because it clearly identifies the symmetries which lead to the topological order in fermionic models and offers a microscopic understanding of the emergence of topological order. Our work is also a concrete step toward finding a Landau like theory for topological order.

	
	
\acknowledgments{Kauffman’s work was supported by the 
	Laboratory of Topology and Dynamics, 
	Novosibirsk State University 
	(contract no. 14.Y26.31.0025 
	with the Ministry of Education and Science 
	of the Russian Federation.)}


\begin{thebibliography}{9}


\bibitem{Kauffman2} L. H. Kauffman, Braiding and Majorana Fermions, in ``Topology and Physics",
edited by C. N. Yang, Mo-Lin Ge and Yang-Hui He, pages 81-108, World Scientific Pub. Co. (2019).

\bibitem{Kitaev} A. Kitaev, Phys. Usp. \textbf{44},131(2001).

\bibitem{Greiter} M. Greiter,V. Schnells and R. Thomale, Annals of Physics
\textbf{351},1026(2014).

\bibitem{Ortiz} E. Cobanera, G. Ortiz and Z. Nussinov, Phys. Rev. B. \textbf{87}, 041105 (2013).

\bibitem{Fendley} Paul Fendley, J. Phys. A: Math. Theor. \textbf{49} (2016).

\bibitem{Sachdev} Subir Sachdev,\textit{Quantum Phase Transitions},Cambridge University Press(2011).

\bibitem{Mier} P. Prelovsek, M. Mierzejewski, O. Barisic, J. Herbrych, Annalen der Physik 529, 1600362 (2017)


\bibitem{Fend2} J. Kemp, N.Y. Yao, C.R. Laumann and P. Fendley, J. Stat. Mech. 063105 (2017)

\bibitem{Nand} R. Nandkishore and D. A. Huse, Ann. Review of Cond. Mat. Phys. 6, 15 (2015)

\bibitem{Altman} E. Altman and R. Vosk, Ann. Review of Cond. Mat. Phys. 6, 383 (2015).

\bibitem{Param} S. A. Parameswaran, A. C. Potter and R. Vasseur, Annalen der Physik 529, 1600302 (2017).

\bibitem{Imbrie} J. Z. Imbrie, V. Ros and A. Scardicchio, Annalen der Physik 529, 1600278 (2017)

\bibitem{Rademaker}  L. Rademaker, M. Ortuno and A.M. Somoza, Annalen der Physik 529, 1600322 (2017).

\bibitem{Luitz}  D. J. Luitz, Y. Bar Lev, Annalen der Physik 529, 1600350 (2017).


\bibitem{Monthus18} Cecile Monthus,J. Phys. A: Math. Theor. 51 265303(2018).

\bibitem{Mier} A. Wieckowski, M. M. Maska and M. Mierzejewski, Phys. Rev. Lett. 120, 040504 (2018)

\bibitem{Rukhsan}  Rukhsan Ul Haq, Resonance, Vol. 21, 12,(2016).

\bibitem{Lee}  J. Lee  and F. Wilczek, Phys. Rev.Lett. \textbf{111}, 226402.

\bibitem{Kells2} G. Kells , Phys. Rev. B. \textbf{89}, 075122(2014).

\bibitem{Yao} H. Yao  and S. A. Kivelson,Phys. Rev. Lett. 99, 247203(2007)

\bibitem{Fendley3} P.  Fendley, Phys. Rev. Lett. 90 120402 (2003).

\bibitem{Beri} B. Beri and N. R. Cooper, Phys. Rev. Lett. 109, 156803 (2012).

\bibitem{Bernevig} Bernevig B. A. 2013 Topological Insulators and Toplogical Superconductors (Princeton University Press, Princeton)

\bibitem{Akhmerov} A. R. Akhmerov, Phys. Rev. B 82, 020509 (2010).

\bibitem{Feldman} Yang G. and Feldman D.E. 2014 Phys. Rev B 89, 035136 



\bibitem{Armin} A. Rahmani et al, Phys. Rev. Lett. 115 166401 (2015).

\bibitem{Grover} T. Groveret al, Science 344, 280 (2014).

\bibitem{Zhang2} Qi X.-L. et al, Phys. Rev. Lett. 102, 187001 (2009).

\bibitem{Hsieh} T. H.  Hsieh et al, Phys. Rev. Lett. 117 166802 (2016).

\bibitem{Chamon} Goldstein G. and Chamon C.,Phys. Rev. B 86, 115122 (2012).

\bibitem{Lounesto} Lounesto Pertti,Clifford Algebras and Spinors, Second Edition, Cambridge University(2001).

\bibitem{Snygg} Snygg John,Clifford Algebra: A computational tool for Physicits,Oxford University Press(1997). 

\bibitem{Else} Else D. V. et al 2017 arxiv:1704.08703

\bibitem{Kells2}  Kells G. 2015 Phys. Rev. B 92, 081401 (2015).
\bibitem{Rademaker} Rademaker Louk et al,Annalen der Physik ,Vol. 529, Issue 7(2017). 
\bibitem{Imbrie} Imbrie J. Z. et al 2017 Ann. Phys. (Berlin) 529, No. 7, 1600278 (2017).
\bibitem{Katsura} Katsura H. et al 2015 Phys. Rev. B 92, 115137 
\bibitem{Ioffe} Doucot B., M. V. Feigel'man, Ioffe L. B. and Ioselevich A. S.,Phys. Rev. B 71, 024505(2005). 
\bibitem{Kou} V. Mourik,K. Zuo, S. M. Frolov, S. R. Plissard, E.P.A.M. Bakkers, L.P. Kouwenhuven, Signatures of Majorana fermions in hybrid superconductor-semiconductor devices, arXiv: 1204.2792.
\bibitem{Moore} Moore G. and Read N. 1991 Nucl. Phys. B 360, 362 
\bibitem{Read} Read N. and Green D. 2000 Phys. Rev. B 61, 10267 
\bibitem{Ivanov} Ivanov D. A. 2001 Phys. Rev. Lett. 86, 268 
\bibitem{Mourik} Mourik V. et al 2012 Science 336, 1003 
\bibitem{Ge} Yu Li-Wei and Ge Mo-Lin, Sci. Rep. 5, 8102 (2015).
\bibitem{Kauffman} L. H. Kauffman  and  S. J. Jr. Lomonaco S.,New. J. Phys. 4,73 (2002).
\bibitem{Kauffman1} Kauffman L. H. 2001 Knots and Physics, World Scientifc, Singapore 

\bibitem{Kauffman3} Kauffman L. H. 2016 Knot logic and topological quantum computing with Majorana fermions. In “Logic and algebraic structures in quantum computing and information”, Lecture Notes in Logic, J. Chubb, J. Chubb, Ali Eskandarian, and V. Harizanov, editors, 124 pages Cambridge University Press

	

\end{thebibliography}
\end{document}